\documentclass{optica-article}

\journal{opticajournal} 

\articletype{Research Article}

\usepackage{lineno}

\usepackage{graphicx}
\usepackage{subfigure}
\usepackage{amsmath}
\usepackage{color}
\usepackage{hyperref}

\def\ket#1{\lvert\nobreak#1\nobreak\rangle}
\def\bra#1{\langle\nobreak#1\nobreak\rvert}

\def\adag{\hat{a}^{\dag}{}}
\def\a{\hat{a}}
\def\sz{\hat{\sigma}_z}
\def\sx{\hat{\sigma}_x}
\def\sy{\hat{\sigma}_y}
\def\splus{\hat{\sigma}_+}
\def\sminus{\hat{\sigma}_-}

\def\alphamag{\lvert \alpha \rvert}
\def\tr{\operatorname{tr}}

\newcommand{\op}[1]{\hat{#1}}

\providecommand{\abs}[1]{\lvert#1\rvert}
\providecommand{\normltwo}[1]{\lVert#1\rVert_{2}}
\providecommand{\normsp}[1]{\lVert#1\rVert_{sp}}
\providecommand{\normsp}[1]{\lVert#1\rVert_{tr}}

\begin{document}

\title{Spectral and dynamical validity of the rotating-wave approximation in the quantum and semiclassical Rabi models}

\author{H. F. A. Coleman\authormark{1} and E. K. Twyeffort\authormark{1,$\dag$,*}}

\address{\authormark{1}School of Physics and Astronomy, University of Southampton, Highfield, Southampton, SO17 1BJ, United Kingdom\\
\authormark{$\dag$}E. K. Twyeffort has previously published as E. K. Irish and E. K. Twyeffort Irish.\\
\email{\authormark{*}e.k.twyeffort@soton.ac.uk}} 


\begin{abstract*} 
Ultrastrong coupling (USC) in the quantum Rabi model, characterized by the breakdown of the rotating-wave approximation (RWA) has become a topic of considerable interest and study. This critical reevaluation of the validity of the RWA concludes that the accepted definition of USC in terms of a fixed ratio of coupling to field frequency is inadequate. Connecting an improved spectral validity criterion with the derivation of the semiclassical limit predicts that the dynamical validity of the quantum RWA should be linked to that of the corresponding semiclassical model. This, however, is not supported by numerical calculations of coherent-state dynamics, which unambiguously demonstrate that spectral validity does not imply dynamical validity and reveal surprisingly complicated dependence on coupling and field amplitude.
\end{abstract*}

\section{Introduction}

Over the past 60 years, the quantum Rabi model has become deeply rooted in theoretical physics. The essence of the interaction between light and matter is distilled into its simplest form: a two-level system (TLS) coupled to a single electromagnetic field mode. A semiclassical form of the model that now bears his name was developed by I. I. Rabi to describe the dynamics of a spin in a classical rotating magnetic field~\cite{Rabi1937}. In 1963, Jaynes and Cummings introduced a version of the model in which the field was treated quantum mechanically to describe the ammonia beam maser and carried out a comparison of the quantum and semiclassical theories~\cite{Jaynes1963}. Although they applied the simplifying approximations now known as the rotating-wave approximations (RWAs) in both the quantum and the semiclassical treatments, it is the exactly solvable quantum model within the RWA that has become most commonly termed the `Jaynes-Cummings model'. 

In the decades that followed, the Jaynes-Cummings model became synonymous with quantum optics and particularly with experiments in cavity quantum electrodynamics (QED)~\cite{Shore1993,HarocheRaimond}. With typical ratios of coupling strength $\lambda$ to field frequency $\omega_0$ on the order of $10^{-7}$--$10^{-6}$~\cite{Hood2000,Raimond2001}, cavity QED with natural atoms lies well within the domain of the RWA. Around the turn of the millennium, advances in nanotechnology inspired the development of engineered solid-state devices exhibiting coherent quantum behavior: superconducting circuits~\cite{Nakamura2002,Vion2002}, nanomechanical resonators~\cite{Cleland1996}, and quantum-well intersubband microcavities~\cite{Anappara2009}, to name but a few. These platforms brought coupling strengths and detunings previously undreamt of in quantum optics into experimental reach. Stimulated by the new capabilities, interest in the quantum Rabi model began to branch out rapidly in new directions.

Much research in the last two decades has centered on the regimes of ultrastrong coupling (USC) and deep strong coupling (DSC)~\cite{LeBoite2020,FornDiaz2019,FriskKockum2019}. Ultrastrong coupling is generally considered to begin where the RWA breaks down~\cite{LarsonMavrogordatos,Ciuti2005,FriskKockum2019,FornDiaz2019}\footnote{When defining the ultrastrong and deep strong coupling regimes, the spin and field are usually assumed to be on or near resonance. Allowing for large detuning adds another dimension to the question of defining parameter regimes within the Rabi model. Here, however, we consider only the resonant case.}. This, of course, necessitates defining the limits of validity of the RWA. A convention has arisen in the USC/DSC community that ultrastrong coupling is reached when $\lambda/\omega_0 \gtrsim 0.1$; see, e.g.~\cite{LarsonMavrogordatos,FriskKockum2019,FornDiaz2019}. In a recent review paper, the authors go so far as to state, ``The lower limit [$\lambda/\omega_0 = 0.1$] has been by now well established as the regime where effects related to the counterrotating terms become sizable and, hence, observable.''~\cite{FornDiaz2019}. 

However, as far back as the 1980s, various authors have observed that the dynamics of the TLS population in the full Rabi model differs noticeably from the RWA predictions even for parameters that satisfy the condition $\lambda/\omega_0 \lesssim 0.1$~\cite{Graham1984,Zaheer1988,Seke1994a,Seke1994b,Finney1994,Pereverzev2006,Naderi2011,He2014,Li2018}. More recent work reflects an increasing recognition that the condition for spectral validity of the RWA does not necessarily justify its application for dynamical calculations~\cite{DeZela1997, Angelo2005, Angelo2007, Larson2012, Burgarth2022, Burgarth2024}, especially in scenarios involving extensions or generalisations of the original Rabi model~\cite{Guido2004, Li2018, Jorgensen2022, Nodar2023, Blaha2023}.

In this work we revisit the question of how to assess and quantify the validity of both the quantum RWA (qRWA) and the semiclassical RWA (scRWA), utilising a combination of numerical computations and heuristic arguments. Two definitions of validity, {\em spectral} and {\em dynamical} validity, are distinguished. We show that the scaling of the spectral validity with excitation number predicts that the validity of the qRWA for a coherent field state $\ket{\alpha}$ is determined not by the magnitude of the coupling $\lambda$ itself but by the product $\lambda\abs{\alpha}$. Drawing on the recently developed mathematical formalism for deriving the semiclassical Rabi model as a limiting case of the quantum Hamiltonian expressed in a displaced Fock basis~\cite{TwyIrish2022}, we argue that the spectral validity condition suggests that the qRWA validity is linked to the validity of the corresponding scRWA. This is tested by carrying out extensive numerical solutions of the system dynamics for various coupling strengths and state amplitudes. Examining several measures of dynamical validity clearly shows that spectral validity does not adequately predict the accuracy of the qRWA for calculations of dynamics. The dynamical validity, in fact, displays complicated dependence on $\lambda$ and $\abs{\alpha}$ individually. We also illustrate numerically the convergence of quantum to semiclassical dynamics and show that non-RWA behavior can be obtained in the semiclassical limit even for values of $\lambda/\omega_0 ~ 10^{-4}$.

To establish the concepts and notation, Sec.~\ref{sec:derivs} begins by briefly reviewing the quantum and semiclassical Rabi models and the derivation of their respective RWA equivalents, together with an outine of the semiclassical limiting procedure of Ref.~\cite{TwyIrish2022}. Various approaches to defining the limits of the spectral validity of the RWA are discussed in Sec.~\ref{sec:spectral} and shown to concur in their predictions of the scaling behavior with excitation number. Sec.~\ref{sec:dynamical} takes up the question of how to assess the dynamical validity of both the quantum and semiclassical RWAs. Convergence of the quantum dynamics and qRWA validity to the semiclassical limit is discussed in Sec.~\ref{sec:converge}. Conclusions, open questions, and outlook for future work are discussed in Sec.~\ref{sec:conclusions}.

\section{Quantum and semiclassical RWAs}\label{sec:derivs}

The quantum Rabi Hamiltonian is given by
\begin{equation}
    \op{H}_q = \frac{\Omega}{2} \sz + \omega_0 \adag \a + \lambda (\adag + \a) \sx ,
\label{Hq}
\end{equation}
where $\Omega$ is the frequency of the two-level system or spin, $\omega_0$ is the frequency of the field, and $\lambda$ is the coupling strength between them. Within the rotating-wave approximation, the Hamiltonian becomes
\begin{equation}
    \op{H}_q^{\rm{RWA}} = \frac{\Omega}{2} \sz + \omega_0 \adag \a + \lambda (\adag \sminus + \a \splus) ,
\label{HqRWA}
\end{equation}
where the spin raising and lowering operators are defined as $\op{\sigma}_{\pm} = \sx \pm i \sy$. This is often derived by transforming $\op{H}_q$ to an interaction picture with respect to $\op{H}_0 = \frac{\Omega}{2} \sz + \omega_0 \adag \a$:
\begin{equation}
    \op{H}_q^I = \lambda\Big[e^{i(\omega_0 - \Omega)t}\adag \sminus + e^{-i(\omega_0 - \Omega)t}\a \splus\Big] + \lambda\Big[e^{i(\omega_0 + \Omega)t}\adag \splus + e^{-i(\omega_0 + \Omega)t}\a \sminus\Big] .
\end{equation}
The terms in the first and second sets of brackets are known as the `co-rotating' and `counter-rotating' terms, respectively. Near resonance ($\Omega \approx \omega_0$), the co-rotating terms have a slow time dependence, while the counter-rotating terms oscillate rapidly. On time scales determined by the spin and field frequencies, these fast oscillations tend to average out and hence have only a small effect on the dynamics. The qRWA consists of neglecting the counter-rotating terms. Alternatively, the qRWA may be viewed as the zeroth-order correction in degenerate perturbation theory with $\lambda$ as the perturbation parameter. The degeneracies occur between $\ket{+z}\ket{n}$ and $\ket{-z}\ket{n+1}$, where $\sz \ket{\pm z} = \pm{\ket{\pm z}}$ and $\adag \a \ket{n} = n \ket{n}$ ($n = 0,1,2,\dots$) denote the eigenstates of the bare Hamiltonians for the spin and field, respectively. Therefore, the qRWA is expected to hold when the spin and field are near resonance ($\abs{\omega_0 - \Omega} \ll \omega_0,\Omega$) and weakly coupled ($\lambda \ll \omega_0, \Omega$). The resulting Hamiltonian is easily solved; for $\Omega = \omega_0$, the eigenstates and energies are given by
\begin{align}
    \ket{\psi_{n,\pm}^{\rm RWA}} &= \frac{1}{\sqrt{2}}(\ket{n,+z} - \ket{n+1,-z}) ,\\
    E_{n,\pm}^{\rm RWA} &= (n+1) \omega_0 \pm \lambda \sqrt{n+1} .
\end{align}

In the semiclassical version of the Rabi model, the field is represented as a classical time-dependent driving term with frequency $\omega_0$ and amplitude $A$, so the system Hamiltonian becomes
\begin{equation}
    \op{H}_{sc} = \frac{\Omega}{2} \sz + 2 A \cos{(\omega_0 t)} \sx .
\label{Hsc}
\end{equation}
Transforming to an interaction picture, this time with respect to the spin term alone, the Hamiltonian may be written as
\begin{equation}
    \op{H}_{sc}^I = A \Big[e^{i(\omega_0 - \Omega) t} \sminus + e^{-i(\omega_0 - \Omega) t} \splus \Big] + A \Big[e^{i(\omega_0 + \Omega) t} \splus + e^{-i(\omega_0 + \Omega) t} \sminus \Big].
\end{equation}
Co-rotating and counter-rotating terms may be identified as before. Neglecting the latter and transforming back to the original frame, the semiclassical RWA Hamiltonian is obtained:
\begin{equation}
    \op{H}_{sc}^{\rm{RWA}} = \frac{\Omega}{2} \sz + A (e^{i\omega_0 t} \sminus + e^{-i\omega_0 t} \splus).
\label{HscRWA}
\end{equation}
The scRWA is valid under conditions of near-resonance and weak driving ($A \ll \omega_0, \Omega$).

Recent work has established that the semiclassical Rabi model can be rigorously derived from the quantum Hamiltonian \cite{TwyIrish2022}. The quantum Hamiltonian is first transformed to a rotating frame with respect to the field term $\omega_0 \adag \a$, then expressed in the orthonormal basis of displaced field states $\ket{\alpha, n} = \op{D}(\alpha) \ket{n}$. By taking the joint limit $\lambda/\omega_0 \to 0$, $\abs{\alpha} \to \infty$ while keeping $\lambda \abs{\alpha}$  constant, 
\begin{equation}
    \op{H}_q \to \op{H}_{sc} \otimes \op{I}_f ,
\label{H_limit}
\end{equation}
where $\hat{H}_{sc}$ is given by Eq.~\eqref{Hsc}. The quantum Hamiltonian separates into a tensor product of the semiclassical Hamiltonian for the spin and the identity operator $\op{I}_f$ for the quantum field, with the product $\lambda \abs{\alpha}$ identified as the semiclassical drive amplitude $A$. Noting that the state $\ket{\alpha, 0}$ is just the standard coherent state $\ket{\alpha}$, this derivation demonstrates that the time evolution of a spin under the Rabi Hamiltonian with the field initialised in a coherent state reduces to the semiclassical spin dynamics in the limit of small coupling and large field amplitude. 

Despite involving a small coupling limit, this semiclassical limiting procedure does not necessarily lead to RWA dynamics. In Ref.~\cite{TwyIrish2022} it is proven that the same procedure can be applied in a polaron-transformed basis often used for studying USC/DSC regimes, which leads to the well-known Bessel function dependence on $A$ for strong off-resonant driving in the semiclassical limit~\cite{Shirley1965,Pegg1973,Lu2012,Ashhab2007}. 

Throughout the rest of this paper, we consider the case of exact resonance, with $\omega_0 = \Omega = 1$. The coherent state amplitude $\alpha$ is taken to be real without loss of generality.

\section{Spectral validity}\label{sec:spectral}

\begin{figure}
    \centering
    \includegraphics[width=1\linewidth]{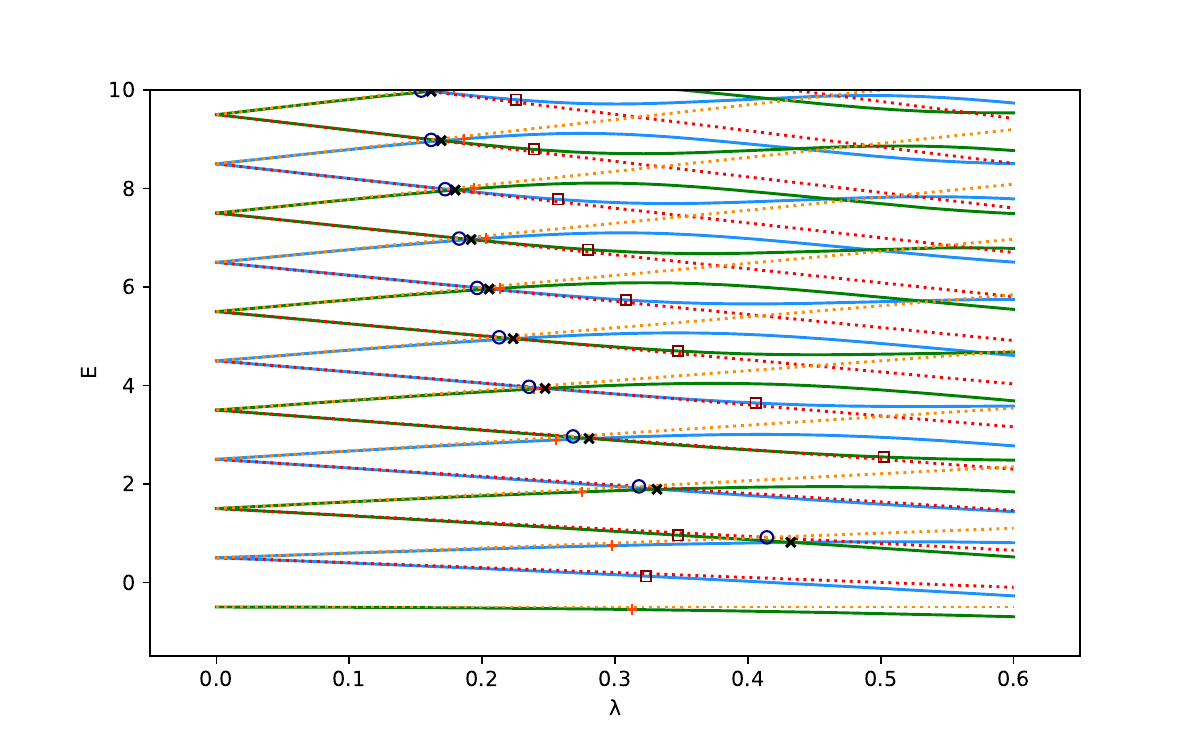}
    \caption{Numerically determined energy spectra for the quantum Rabi model. Solid lines correspond to the full model, with states of opposite parities indicated as blue and green. RWA energies are shown as dotted lines; red and orange denote the upper and lower states in each pair. The individual points indicate different definitions for the point $\lambda_c$ where the RWA breaks down: $\lambda_{c}^{\rm RWA}$ ($\circ$), $\lambda_c^{\rm pUSC}$ ($\times$), $\lambda_s^{n,+}$ ($+$), $\lambda_s^{n,-}$ ($\square$). }
    \label{spectrum}
\end{figure}

While the typical statement of validity for the RWA as $\lambda/\omega_0 \lesssim 0.1$ can be useful as a rule of thumb, a comparison of the energy spectra of the quantum Rabi model with and without the qRWA, plotted in Fig.~\ref{spectrum}, shows that it is reasonable only for the lowest few energy levels. (An alternative argument leading to the same criterion is made in Ref.~\cite{Wolf2013}.) The range of coupling strengths over which the RWA holds decreases as the levels go up in energy. For physical situations involving higher energy levels, this dependence needs to be taken into account when assessing the validity of the RWA. 

Examining the two spectra suggests that the RWA roughly holds for coupling strengths up to the point where the two states $\psi_{n,+}^{\rm RWA}$ and $\psi_{n+1,-}^{\rm RWA}$ cross, which we denote $\lambda_{c}^{\rm RWA}$~\cite{Feranchuk1996, TwyIrish2005, TwyIrish2007}. This crossing point can easily be expressed analytically:
\begin{equation}
    \frac{\lambda_{c}^{\rm RWA}}{\omega_0} = \frac{1}{\sqrt{n} + \sqrt{n+1}} \approx \frac{1}{2\sqrt{n}},
\end{equation}
where the approximation holds in the limit of large $n$. Alternatively, the left-hand expression may be derived by calculating the perturbative corrections from the counter-rotating terms~\cite{DeBernardis2024}.

The degeneracies between energy levels in the full Rabi model, known as Juddian points or exceptional eigenvalues, are more difficult to determine. In principle they may be found exactly using recursion relations~\cite{Braak2011}; in practice this requires numerical evaluation. An approximation was obtained in \cite{Rossatto2017}, which the authors use to define the upper boundary of the `perturbative ultrastrong coupling' (pUSC) regime. Their expression is
\begin{equation}
    \frac{\lambda_c^{\rm pUSC}}{\omega_0} = \frac{1}{\sqrt{2(2n+1)}} .
\end{equation}
As expected from the spectral plots, this converges rapidly to the RWA crossing point as $n$ increases.

\begin{figure}
    \centering
    \includegraphics[width=1\linewidth]{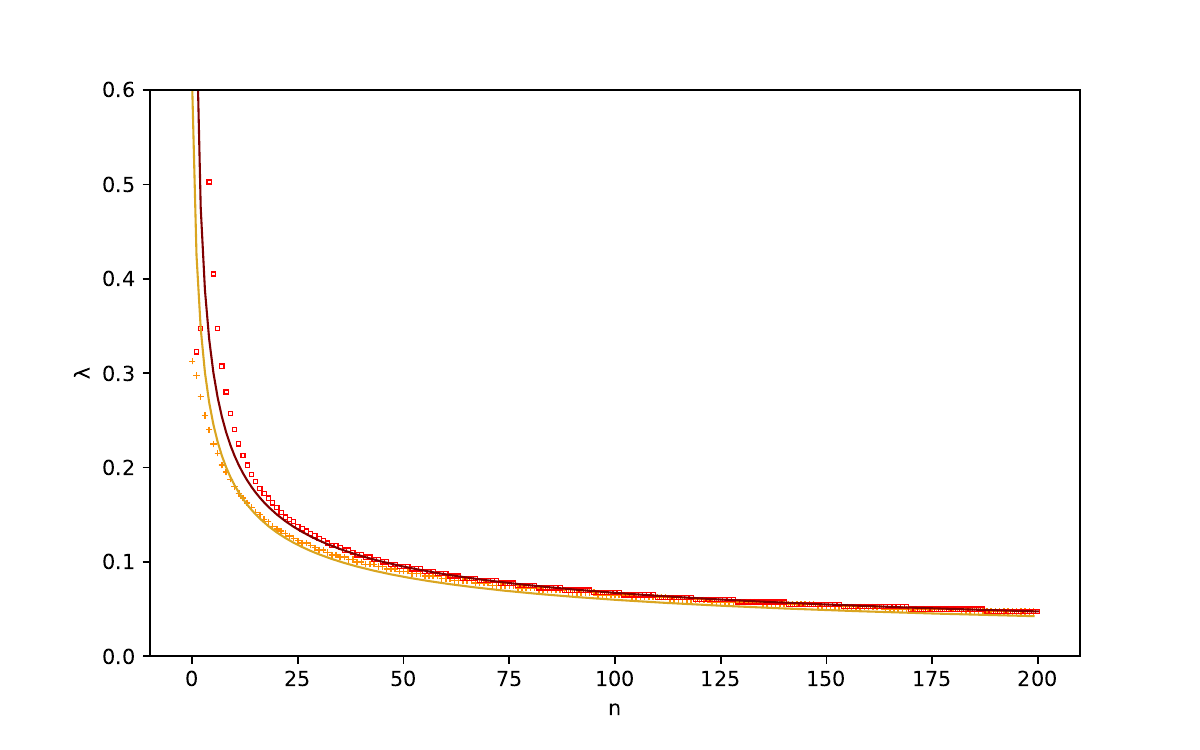}
    \caption{Numerically determined splitting points $\lambda_s^{n,-}$ (red squares) and $\lambda_s^{n,+}$ (orange crosses); the colors and point styles are the same as in Fig.~\ref{spectrum}. The solid curves are fits with functional dependence $1/\sqrt{n}$ (red) and $1/\sqrt{n+1}$ (orange).}
    \label{splittings}
\end{figure}

To complement these two analytical approximations, we numerically evaluate the differences between the quantum and RWA spectra. The `splitting point' $\lambda_s$ for each eigenvalue is defined as the point where the difference between the full quantum model and its RWA counterpart exceeds a predefined threshold $\delta_\lambda$. (In the plots shown, $\delta_\lambda/\omega_0 = 0.05$; the conclusions remain similar for any sensible choice.) Curiously, this analysis reveals a discrepancy in the $\lambda_s$ values between the two levels involved in the first crossing. For the levels with an initial downward slope, which in the RWA are given by $\ket{\psi_{n,-}^{\rm RWA}}$, the splitting point $\lambda_s^{n,-}$ is consistently larger than the values $\lambda_s^{n,+}$ and $\lambda_s^{n+1,+}$ associated with the adjacent upward-sloping levels. This may be seen from the points plotted in Fig.~\ref{spectrum}. In Fig.~\ref{splittings} the splitting points are plotted as a function of $n$, with $\lambda_s^{n,\pm}$ indicated separately. The solid lines are fits of the form $1/\sqrt{n}$ (lower states) and $1/\sqrt{n+1}$ (upper states); the difference is due to how the excitation number $n$ is defined. The discrepancy in $\lambda_s$ between the two sets of states is clearly visible. Understanding the origin of this effect is beyond the scope of the present work, but we note it as an interesting avenue for future exploration. For our purposes, the key point is that the fits capture the scaling behavior of $\lambda_s$ for large $n$. 

All of these approaches agree that the coupling $\lambda_c$ at which the RWA breaks down scales as $\lambda_c/\omega_0 \approx 1/(2 \sqrt{n})$ for large values of $n$, with the differences becoming small even for $n \sim 10$. A coherent state $\ket{\alpha}$ of the field contains a distribution of Fock states, with average photon number $\bar{n} = \abs{\alpha}^2$. For large $\alpha$, the distribution is sharply peaked around this average. Provided that $\lambda/\omega_0 \ll 1/(2 \sqrt{\bar{n}})$, all the eigenstates with significant probability amplitude in the coherent state will fall below their individual values of $\lambda_c$. Spectral validity, then, predicts that the qRWA will hold for a coherent state $\ket{\alpha}$ if $\lambda/\omega_0 \ll 1/(2 \abs{\alpha})$. In turn, this condition may be connected with the semiclassical model by noting that $\lambda \abs{\alpha} \equiv A$ in the derivation of the semiclassical limit from the quantum model. This implies that requiring the qRWA to hold for the eigenstates that make up the initial coherent state automatically satisfies the condition $A/\omega_0 \ll 1$; conversely, if the condition $A/\omega_0 \ll 1$ for the validity of the scRWA in the corresponding semiclassical case is met, the qRWA will be valid for any coherent state $\ket{\alpha}$. Put another way, spectral validity of the qRWA predicts that the validity of the RWA for a coherent state of the field should depend only on the product $\lambda \abs{\alpha}$. In the following section, we examine whether this prediction proves correct.

\section{Dynamical validity}\label{sec:dynamical}

\begin{figure}
    \centering
    \includegraphics[width=1\linewidth]{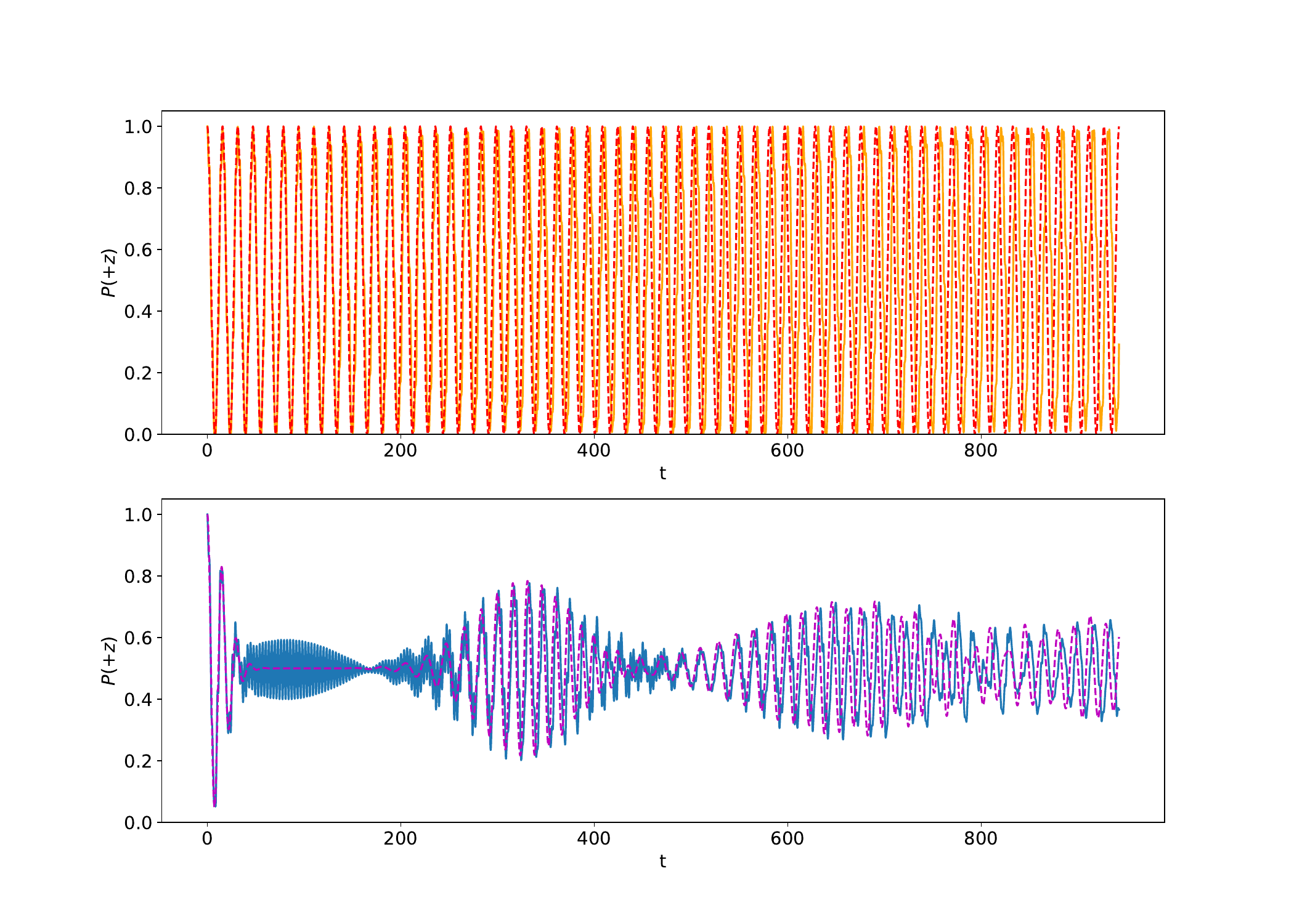}
    \caption{Dynamics of the excited-state spin population in the full Rabi model (solid lines) and RWA (dashed lines). The upper panel shows the semiclassical case and the lower panel the quantum case. The spin is initially in the excited state $\ket{+z}$. The semiclassical field amplitude $A/\omega_0 = 0.2$; the quantum field begins in a coherent state $\ket{\alpha}$ with $\alpha = \sqrt{10}$ and the coupling is $\lambda/\omega_0 = 0.2/\sqrt{10} \approx 0.06$.}
    \label{pop_evol}
\end{figure}

Although it is not so widely recognized in the field, reservations about the validity of the RWA for dynamical calculations date back to long before the advent of the modern USC/DSC terminology. In one influential example, Zaheer and Zubairy showed substantial deviations from RWA dynamics for exact resonance with coupling $\lambda/\omega_0 = 0.2/\sqrt{10} \approx 0.06$~\cite{Zaheer1988}. The initial state of the field was taken to be a coherent state of amplitude $\alphamag = \sqrt{10}$, corresponding to $A/\omega_0 = 0.2$. Figure~\ref{pop_evol} shows the time evolution of the excited-state population under the full and RWA Hamiltonians, in both the semiclassical and quantum versions of the Rabi model, calculated by numerical solution of the Schr{\"o}dinger equation. The parameters are the same as those in Ref.~\cite{Zaheer1988}, except that the maximum time has been extended. Significant deviations between the full and RWA quantum dynamics can be seen, particularly the fast oscillations within the `quiescent region' following the initial collapse of Rabi oscillations. More recent work has also highlighted the failure of the RWA for dynamics even when the standard validity conditions are met~\cite{DeZela1997, Angelo2005, Angelo2007, Larson2012, Burgarth2022, Burgarth2024,Guido2004, Li2018, Jorgensen2022, Nodar2023, Blaha2023}. This leads us to consider the {\em dynamical validity} of the approximation.

Defining an appropriate figure of merit for quantifying the dynamical validity is not straightforward. The dynamics of both spin and field are complicated, with multiple competing timescales. Conclusions regarding the applicability of the RWA have often been based on qualitative examinations of numerically evaluated spin dynamics~\cite{Graham1984,Zaheer1988,Seke1994a,Seke1994b}. Furthermore, it is generally acknowledged that the goodness of the approximation depends on the timescale of interest. The spectral validity condition $\lambda \sqrt{\bar{n}} \ll \omega_0$ can also be derived from time-dependent perturbation theory~\cite{Puri}, which is suggestive that it may only hold over short times.

Recently, some efforts have been made to mathematically bound the dynamical validity of the RWA. In the semiclassical case, Angelo and Wreszinski considered the difference between a state $\psi(t)$ evolving under the full Rabi Hamiltonian and the corresponding state $\psi_{\text{RWA}}(t)$ in the RWA, taking as their measure the norm of the difference between the state vectors and showing that
\begin{equation}
\normltwo{\psi(t) - \psi_{\text{RWA}}(t)} = \mathcal{O}(2A\tau) \qquad 0 \leq t \leq \mathcal{O}\left(1/(2A\tau)\right),
\label{AW_bound}
\end{equation}
where $\tau = 2\pi/\omega_0$ and we use $\normltwo{\cdot}$ to denote the Euclidean or $\ell^2$ norm in the spinor space~\cite{Angelo2005,Angelo2007}. Burgarth and coworkers have taken a different approach, based on bounding the approximation in terms of propagators~\cite{Burgarth2022, Burgarth2024}\footnote{Here we have simply translated Angelo and Wreszinski's result into our notation. It should be noted that $1/(2A\tau)$ does not have units of time; however, as the bound has been stated only up to a constant, this makes no significant difference.}. For the semiclassical case, their estimate for the error of the RWA in a time period $T$ becomes, for exact resonance,
\begin{equation}
    \normsp{U(t) - U_{\text{RWA}}(t)} \leq \frac{\abs{A}}{\omega_0} (1 + 4AT) \qquad 0 \leq t \leq T ,
    \label{Bur_sc_bound}
\end{equation}
where $U(t)$ and $U_{\text{RWA}}(t)$ denote the propagators corresponding to the full and RWA Hamiltonians, respectively, and $\normsp{\cdot}$ denotes the spectral norm for operators~\cite{Burgarth2022}. Generalizing these methods to the quantum case, Ref.~\cite{Burgarth2024} provides both upper and lower bounds on the error of the approximation in terms of the state vectors. However, the lower bound is restricted to very short times, $t \leq \pi/\omega_0$, rendering it of limited use in practical applications. The upper bound, again for exact resonance, is stated as
\begin{equation}
    \normltwo{(U(t) -  U_{\text{RWA}}(t)) \Psi} \leq \frac{\lambda}{\omega_0} \left[ \normltwo{(\adag \a + 2)^{1/2} \Psi} + \abs{t} \left(3 \lambda \normltwo{((\adag \a + 2)(\adag \a + 3))^{1/2} \Psi} \right) \right]
    \label{Bur_q_bound}
\end{equation}
for all times $t$, where $\Psi$ is an arbitrary initial state of the coupled quantum system. Note that the left-hand side of Eq.~\eqref{Bur_q_bound} is equivalent to the norm difference between the state vectors, the same measure used in Eq.~\eqref{AW_bound}.

\begin{figure}
    \centering
    \includegraphics[width=1\linewidth]{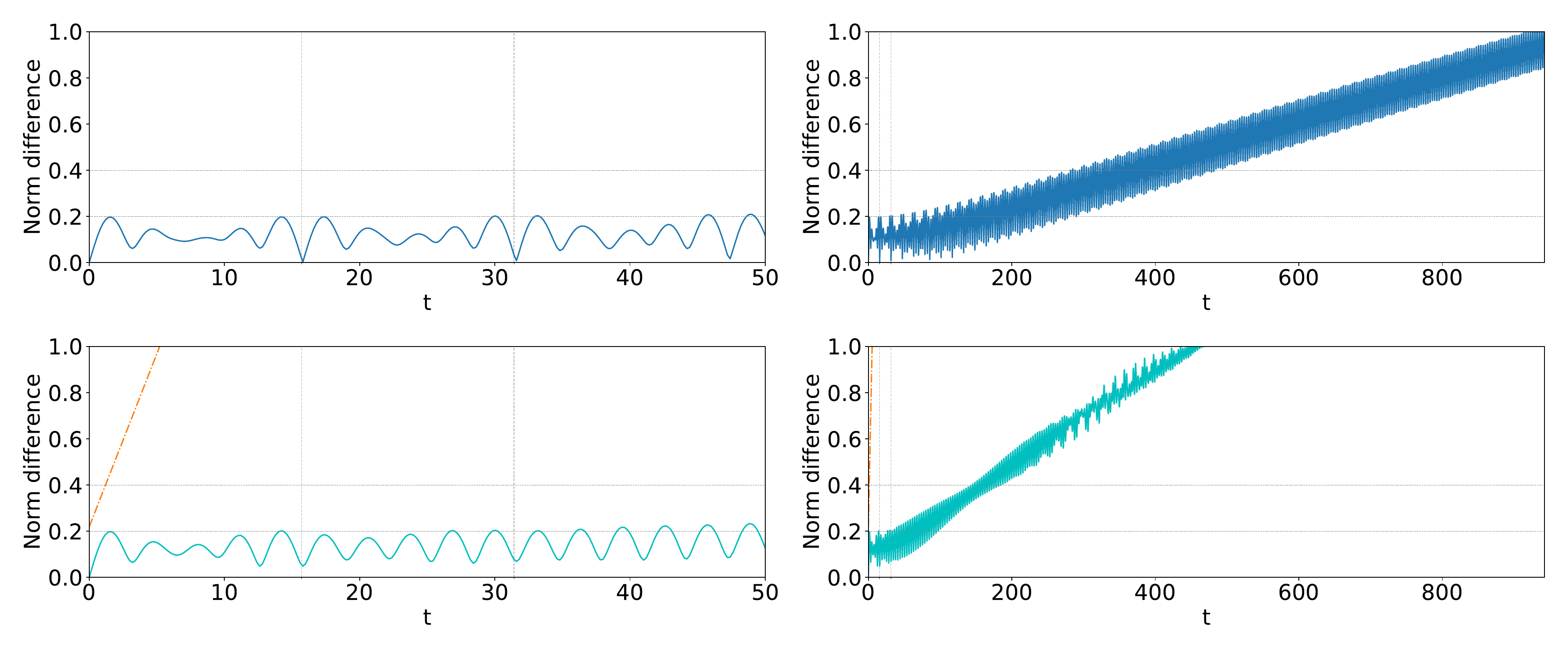}
    \caption{Numerical evaluation of the norm of the difference between full and RWA states, $\normltwo{\psi(t) - \psi_{\text{RWA}}(t)}$, in the semiclassical (upper panels) and quantum (lower panels) models. Two different timescales are shown. The dot-dashed orange lines indicate the bound of Eq.~\eqref{Bur_q_bound}~\cite{Burgarth2024}. Vertical dashed lines indicate $t = \{\tau_R, 2\tau_R\}$, where $\tau_R = \pi/A$ is the Rabi period in the semiclassical RWA; horizontal dashed lines correspond to $\{A, 2A\}$.}
    \label{normdiff}
\end{figure}

\begin{figure}
    \centering
    \includegraphics[width=1\linewidth]{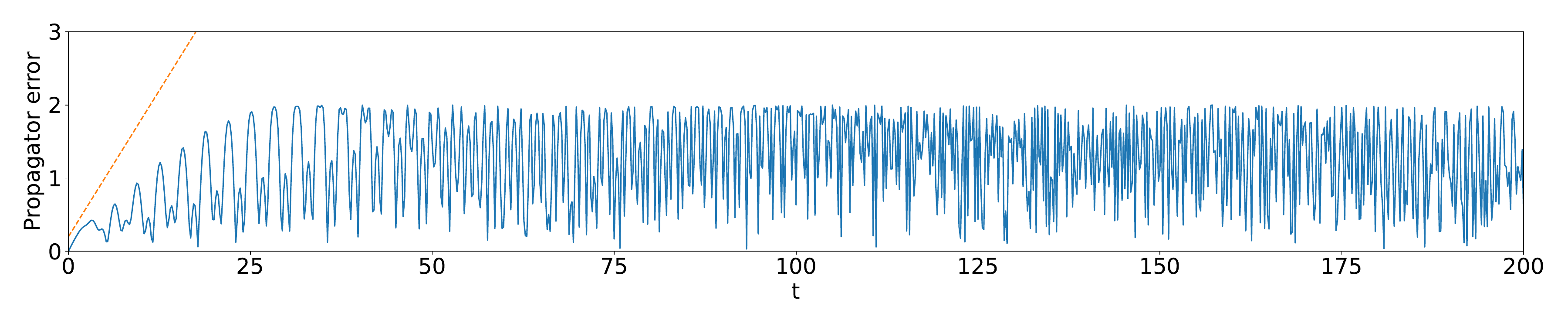}
    \caption{Numerical evaluation of the difference between the propagators $\normltwo{(U(t) -  U_{\text{RWA}}(t))}$ for the semiclassical model, together with the bound of Eq.~\eqref{Bur_sc_bound} (orange dashed line) \cite{Burgarth2022}.}
    \label{prop_err}
\end{figure}

However, these bounds fail to capture the full picture of the dynamical validity of the RWA, either semiclassical or quantum. A numerical evaluation of the norm of the difference between state vectors described by Eqs.~\eqref{AW_bound} and \eqref{Bur_q_bound} is shown in Fig.~\ref{normdiff}, for the same set of parameters as Fig.~\ref{pop_evol}. The semiclassical results for both short and long times are shown in the top panels. Vertical dashed lines indicate $t = \{\tau_R, 2\tau_R\}$, where $\tau_R = \pi/A$ is the Rabi period in the semiclassical RWA; horizontal dashed lines correspond to $\{A, 2A\}$. These lines provide a guide for comparing the numerical results in Fig.~\ref{pop_evol} with the relation given in Eq.~\eqref{AW_bound}. While the results indicate that Eq.~\eqref{AW_bound} is satisfied, the relation is of limited usefulness in describing the actual behavior of $\normltwo{\psi(t) - \psi_{\text{RWA}}(t)}$. The corresponding results for the fully quantum model are plotted in the lower panels of Fig.~\ref{pop_evol}, with the dash-dotted line indicating the upper bound given in Eq.~\eqref{Bur_q_bound}. While the bound holds, it is far from tight. A similar conclusion follows for the bound on the difference between the semiclassical propagators given by Eq.~\eqref{Bur_sc_bound}, as seen from Fig.~\ref{prop_err}.  

A more serious issue with the bounds of Eqs.~\eqref{AW_bound} and \eqref{Bur_q_bound} is the choice of the $\ell^2$ norm of the difference between state vectors as the measure of validity. This quantity can be expressed as
\begin{equation}
    \normltwo{\psi_1 - \psi_2} = \sqrt{2 - \langle \psi_1 \vert \psi_2 \rangle - \langle \psi_2 \vert \psi_1\rangle} = \sqrt{2\bigl[1 - \operatorname{Re}\bigl(\langle \psi_1 \vert \psi_2 \rangle\bigr)\bigr]}.
\end{equation}
The fact that $\normltwo{\psi_1 - \psi_2}$ only depends on the real part of the inner product between the states makes it problematic as a measure. For instance, consider $\psi_2 = e^{i \phi} \psi_1$ with $\phi$ real. Physically, $\psi_1$ and $\psi_2$ represent the same state regardless of the value of $\phi$. The norm of their difference, however, is $\normltwo{\psi_1 - \psi_2} = \sqrt{2(1-\cos \phi)}$, which varies between 0 and 2 depending on $\phi$. While the relative phase between two states does have meaning, the norm of the difference fails to distinguish a phase difference from a difference in the state vectors themselves.

\begin{figure}
    \centering
    \includegraphics[width=1\linewidth]{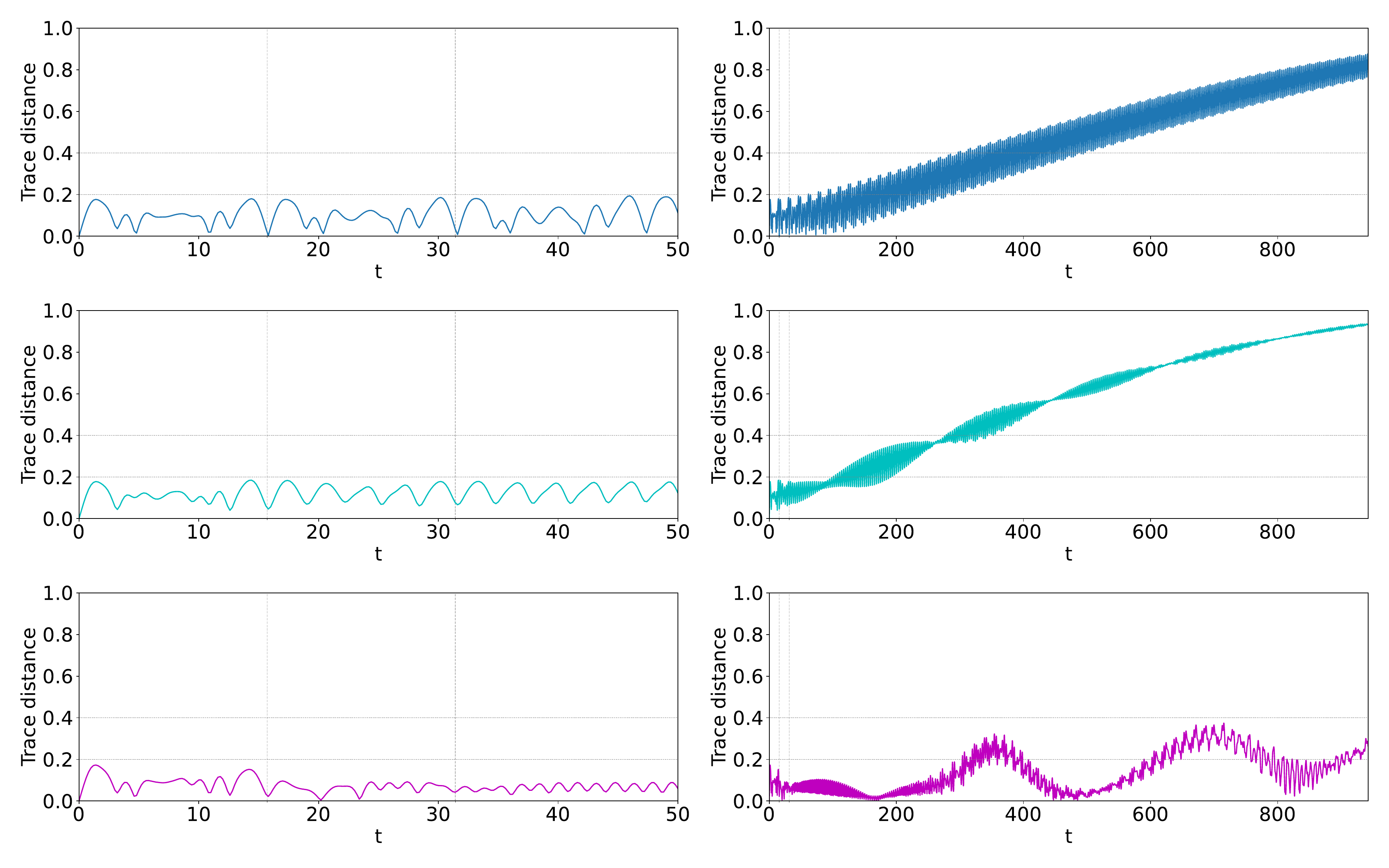}
    \caption{Trace distances between the same initial state evolving under the full and RWA models. The top panels show the semiclassical case. The middle is the trace distance between the full state vectors in the quantum case, while the bottom is the trace distance between the reduced density matrices for the spin. The dashed lines are the same as in Fig.~\ref{normdiff}.}
    \label{tracedist}
\end{figure}

We suggest that a better choice for measuring the divergence between two states is the trace distance between density matrices $\rho_1$ and $\rho_2$,
\begin{equation}
    D(\rho_1, \rho_2) \equiv \tfrac{1}{2} \tr \abs{\rho_1 - \rho_2} ,
\end{equation}
which constitutes a valid metric on quantum states~\cite{NielsenChuang}. The trace distance between state vectors evolving under the full and RWA Hamiltonians in the semiclassical case is plotted in the top panels of Fig.~\ref{tracedist}. Over short timescales, the trace distance closely resembles the norm difference shown in Fig.~\ref{normdiff} (top left). This may be understood by noting that for pure states $\ket{\psi_1}$ and $\ket{\psi_2}$, the trace distance can be expressed as
\begin{equation}
    D(\ket{\psi_1}\bra{\psi_1}, \ket{\psi_2}\bra{\psi_2}) = \sqrt{1 - \abs{\langle \psi_1 \vert \psi_2 \rangle}^2} .
\end{equation}
For two states that differ by a small amount, $\langle \psi_1 \vert \psi_2 \rangle = 1 - \epsilon$, 
\begin{align}
\normltwo{\psi_1 - \psi_2} &= \sqrt{2 \operatorname{Re}(\epsilon)} \\
 D(\ket{\psi_1}\bra{\psi_1}, \ket{\psi_2}\bra{\psi_2}) &= \sqrt{2 \operatorname{Re}(\epsilon)} + \mathcal{O}(\epsilon^{3/2}),
\end{align}
so the norm difference and trace distance approximately agree. Over long times, substantial disagreement emerges; this can already be seen by comparing the top right panels of Fig.~\ref{tracedist} and Fig.~\ref{normdiff}. (Note that the norm difference and trace distance are defined on different intervals: $[0,2]$ and $[0,1]$ respectively.) In the quantum case, the trace distance between the state vectors of the spin--field system, shown in the middle row of Fig.~\ref{tracedist}, again resembles the norm difference [Fig.~\ref{normdiff}, lower panels] for short times, but increases much less rapidly over a longer timescale. Here the agreement between the semiclassical and quantum results is much closer and the differences are more evidently linked to the collapse and revival behavior in the quantum model.

A further argument for choosing the trace distance as a measure is that it can be applied to density matrices as well as state vectors. In experiments and applications, it is often the dynamics of the spin alone that is of interest, rather than the full quantum state of the spin--field system. The trace distance $D(\rho_{s}, \rho_{s}^{\mathrm{RWA}})$, where $\rho_{s}$ and $\rho_{s}^{\mathrm{RWA}}$ are the reduced density matrices of the spin in the full and RWA cases, respectively, is shown in the bottom row of Fig.~\ref{tracedist}. Intriguingly, the behavior in this case differs substantially from that of the semiclassical and quantum state vectors. The trace distance remains small over a much longer timescale, with a structure that shows a clear relationship to the collapse and revival dynamics of the spin population (c.f. Fig.~\ref{pop_evol}).  

\begin{figure}
    \centering
    \includegraphics[width=1\linewidth]{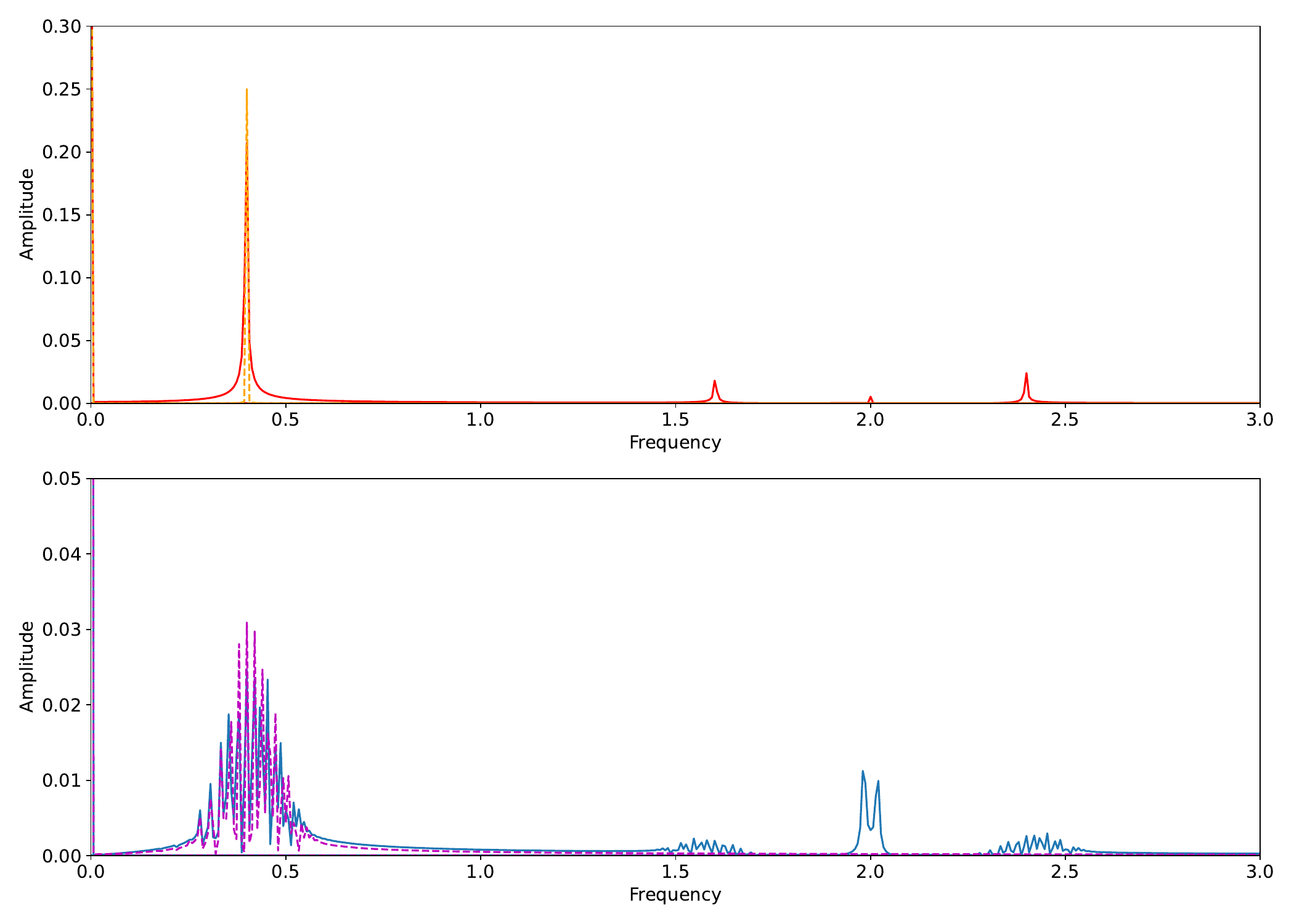}
    \caption{Fourier transforms (FFT) of the spin excited-state probability in the full (solid lines) and RWA (dashed lines) versions of the semiclassical (top) and quantum (bottom) models. Note the differing scales on the $y$-axes.}
    \label{fft}
\end{figure}

\begin{figure}
    \centering
    \includegraphics[width=0.75\linewidth]{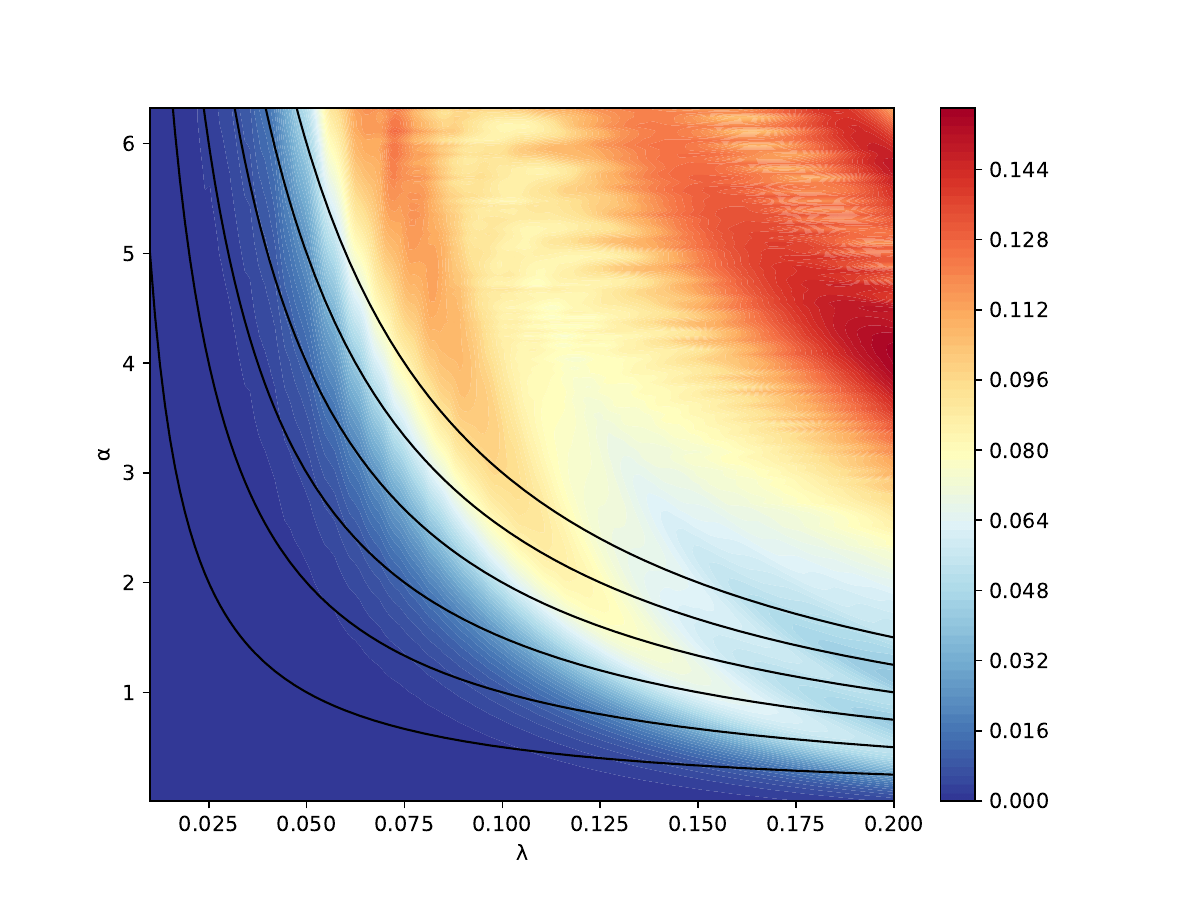}
    \caption{Contour plot of the correlation $1-r_q^2$ between the full and RWA Fourier spectra in the quantum model, as a function of the coupling $\lambda$ and the coherent state amplitude $\alpha$. Solid lines denote constant values of $A = \lambda \alpha = \{0.05,~0.1,~0.15,~0.2,~0.25,~0.3$\}.}
    \label{corr_trev}
\end{figure}

\begin{figure}
    \centering
    \includegraphics[width=1\linewidth]{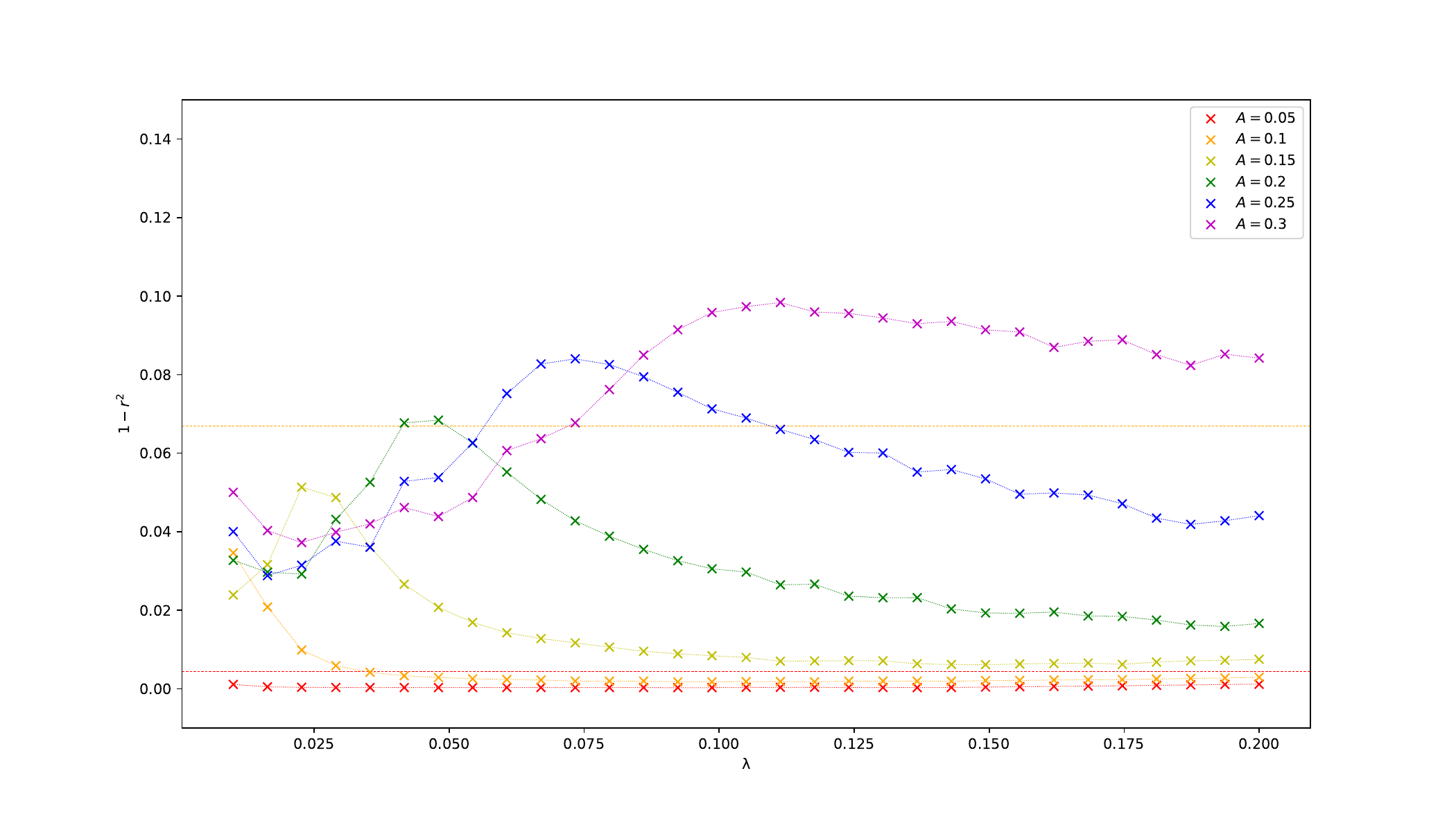}
    \caption{Correlations $1-r_q^2$ as a function of $\lambda$ for constant values of $A$, corresponding to the solid lines in Fig.~\ref{corr_trev}. Dashed horizontal lines indicate the values of the semiclassical correlations; for $A > 0.1$, these lie well above the $y$-scale shown. Dotted lines connecting the points are provided as a guide to the eye.}
    \label{corrs_const_A}
\end{figure}

While examining the time dependence of the trace distance is instructive, it is a cumbersome tool to employ for comparing the validity of the RWA across a range of parameter values. The contributions of the counter-rotating terms are more readily identified in the frequency domain. Figure~\ref{fft} shows the Fourier transforms (FFT) of the spin excited-state population dynamics plotted in Fig.~\ref{pop_evol}. To convert the Fourier spectra into a single figure of merit, we compute the Pearson correlation coefficient $r$ between the full and RWA spectra. Since $r$ is always near 1 for the relevant parameter ranges, we instead plot $1 - r^2$ to highlight small differences from perfect correlation. We refer to this quantity as the correlation. A contour plot demonstrating how the quantum correlation $1 - r_q^2$ depends on $\lambda$ and $\alpha$ is shown in Fig.~\ref{corr_trev}. Another view of the correlations is shown in Fig.~\ref{corrs_const_A}, where both the quantum and semiclassical correlations are plotted as a function of $\lambda$ for several values of $A$. 

According to the spectral validity prediction, the quantum correlation should be constant for a fixed value of $A$. Clearly, this is not borne out by the dynamical calculations. Rather, the dependence of the quantum correlation on $\lambda$ displays a surprisingly complicated structure. This is likely a consequence of basing the calculation on the dynamics of the spin alone. We speculate that the correlation is particularly sensitive to the Bloch-Siegert shift, the effect of which tends to be suppressed by the collapse of the Rabi oscillations. Detailed analysis of the behavior of the correlation is a subject for future work, however. Within the scope of this paper, it suffices to observe that the spectral validity condition is insufficient to determine the goodness of the RWA for calculating spin dynamics with a coherent quantum field.

\section{Convergence to the semiclassical limit}\label{sec:converge}

The observant reader may have noticed that Fig.~\ref{corrs_const_A} does not show the quantum correlation converging toward the semiclassical correlation for small $\lambda$, as would be expected from the semiclassical limiting procedure laid out in Ref.~\cite{TwyIrish2022} and summarised briefly in Sec.~\ref{sec:derivs}. In this section we elucidate the circumstances under which semiclassical convergence is obtained and illustrate this process with numerical results.

With the field initially in a coherent state, the spin dynamics in the qRWA (and the full Rabi model as long as $\lambda$ is not too large) is complicated by the coexistence of multiple timescales. Borrowing the well-known approximate results of Eberly {\em et al.}~\cite{Eberly1980} and assuming exact resonance, we can identify the Rabi period $\tau_R \propto 1/(\lambda \alpha) = 1/A$, the collapse time $\tau_{col} \propto 1/\lambda$, and the revival time $\tau_{rev} \propto \alpha/\lambda$. Recovering semiclassical behavior requires pushing the collapse and revival times to infinity while keeping the Rabi period fixed~\cite{HarocheRaimond,LarsonMavrogordatos}; this is precisely what is achieved by the joint limit $\lambda \to 0$ and $\abs{\alpha} \to \infty$ with $\lambda \abs{\alpha}$ kept constant \cite{TwyIrish2022}. The key observation is that the appearance of semiclassical behavior depends on the timescale over which the dynamics is measured.  

The aim of Figs.~\ref{corr_trev} and \ref{corrs_const_A} was to compare the validity of the qRWA, as measured by the correlation, between different instances of the quantum dynamics. To do this, it is necessary to resolve all the frequencies in the quantum spectra, which requires computing the dynamics over long timescales to capture the full collapse and revival behavior (here, $3 \tau_{rev}$). It follows that neither the dynamics, the spectra, or the correlations will appear semiclassical in nature.

\begin{figure}[tbp]
    \centering
    \includegraphics[width=1\linewidth]{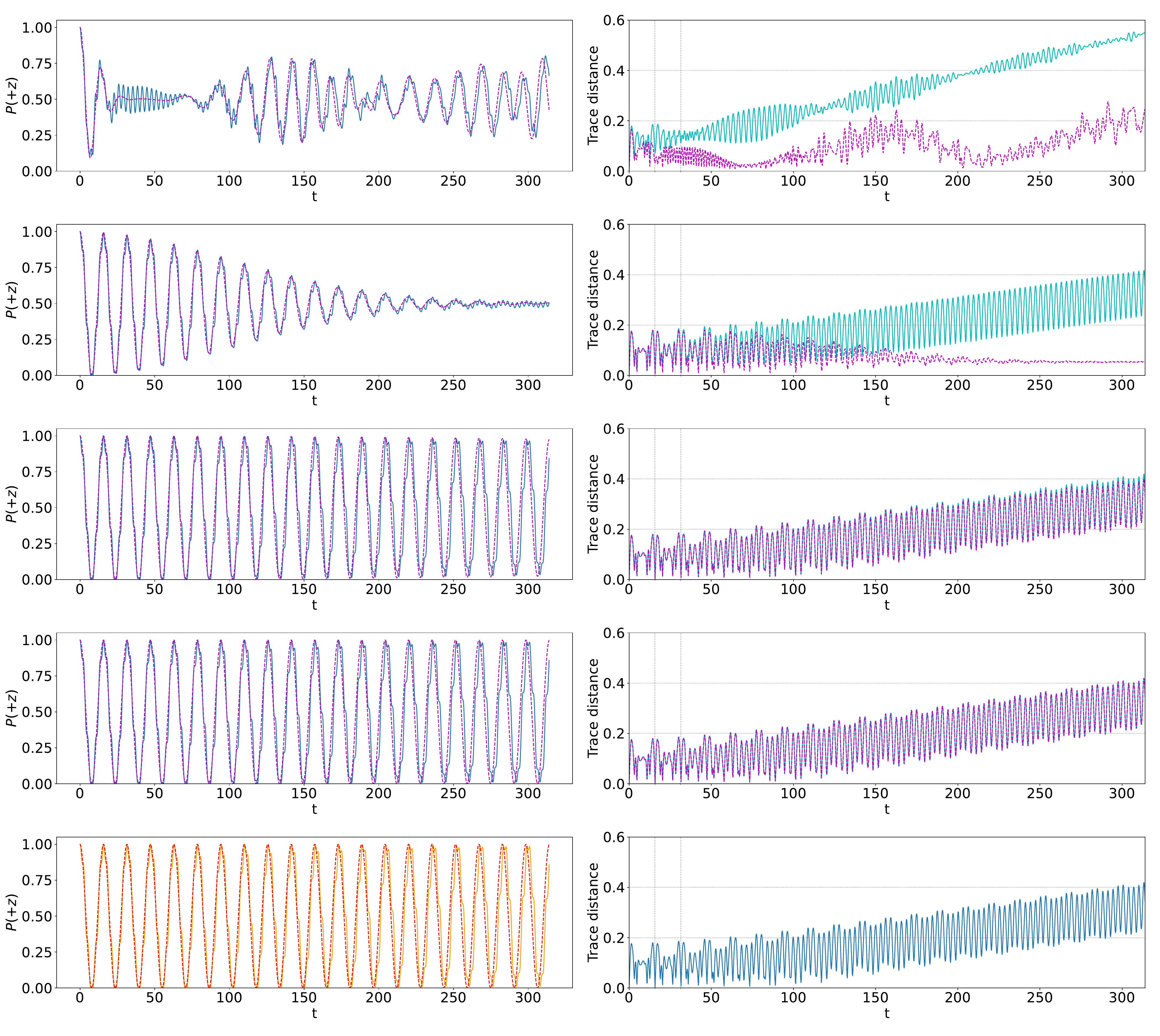}
    \caption{Convergence of the quantum evolution to the semiclassical limit for $A = \lambda \alpha = 0.2$. From top to bottom: quantum with $\lambda = 10^{-1},10^{-2},10^{-3},10^{-4}$; semiclassical. Left: Dynamics of the excited-state spin population in the full model (solid lines) and RWA (dashed lines). Right: Trace distances between the full and RWA states. Solid curves correspond to the trace distance between the state vectors (solid curves); the dashed magenta curve corresponds to the trace distance between the reduced spin density matrices for the quantum case.}
    \label{sc_converge}
\end{figure}

\begin{figure}
    \centering
    \includegraphics[width=1\linewidth]{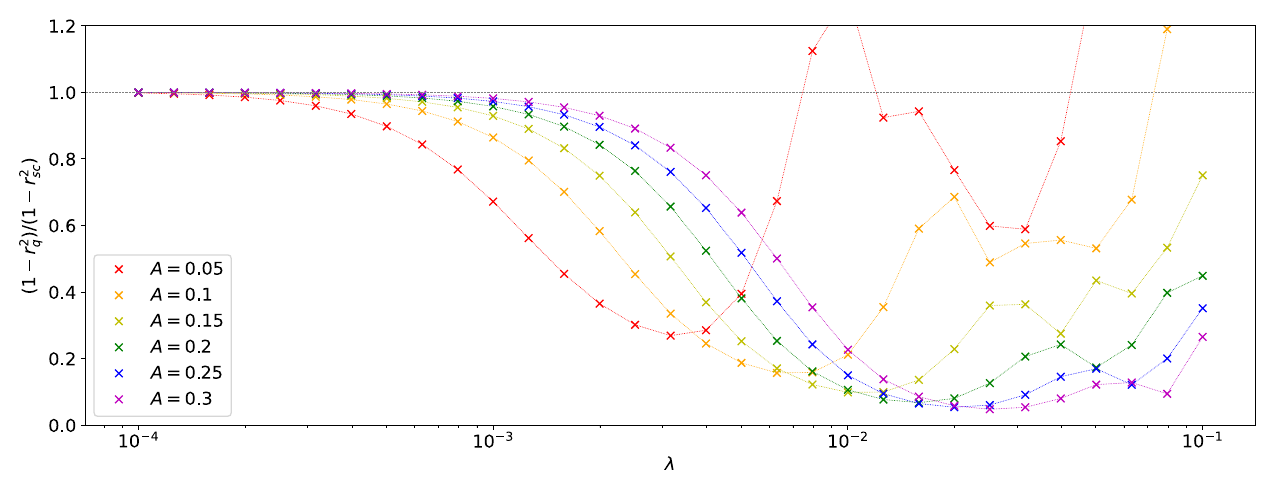}
    \caption{Correlations $1-r_q^2$ for the quantum dynamics over 20 Rabi periods, as a function of the quantum coupling $\lambda$. For each value of $A$, the points have been normalised to the value of the corresponding semiclassical correlation, so the grey dashed line at 1 indicates where $1-r_q^2 = 1-r_{sc}^2$. Dotted lines connecting the points are provided as a guide to the eye.}
    \label{corr_coeffs_sc_converge}
\end{figure}

To see the emergence of semiclassical behavior, the spin dynamics is instead calculated over a timescale corresponding to a fixed number of Rabi periods. Figure \ref{sc_converge} illustrates how the quantum dynamics converges to the semiclassical dynamics as $\lambda/\omega_0$ is decreased from $10^{-1}$ to $10^{-4}$, with $A$ fixed at 0.2. Plots of the trace distance between the full and RWA states are also shown. The correlation also provides a useful figure of merit for studying the semiclassical convergence. A plot of the correlation as a function of $\lambda$ for different values of $A = \lambda \alpha$ is shown in Fig.~\ref{corr_coeffs_sc_converge}. In order to compare the rate of convergence for different $A$, the results have been normalised to the corresponding semiclassical correlations. This figure demonstrates that it is necesssary to go out to values of $\lambda/\omega_0 \sim 10^{-4}$ to see agreement between the quantum and semiclassical dynamics, as suggested by Fig.~\ref{sc_converge}. Calculation of the dynamics with such large coherent states ($\bar{n} = \abs{\alpha}^2 \sim 10^6$) is made computationally tractable by working in the displaced Fock basis described in Ref.~\cite{TwyIrish2022}, which maps any initial coherent state to $n = 0$. As $\alpha$ increases, the rate at which the field state changes becomes slower relative to the Rabi frequency, so for a moderate number of Rabi oscillations the dynamics may be accurately simulated with reasonable matrix sizes.

\clearpage

The plots in Figs.~\ref{sc_converge} and \ref{corr_coeffs_sc_converge} both demonstrate the convergence of the quantum dynamics to the semiclassical result as $\lambda/\omega_0 \to \infty$. Surprisingly, even for values of $\lambda/\omega_0$ three orders of magnitude smaller than the standard USC boundary, the qRWA does not necessarily hold. Rather, the degree of validity of the qRWA converges to that of the corresponding scRWA. Applying the semiclassical limiting procedure to the full quantum Hamiltonian $H_q$ gives back $H_{sc}$, which is only approximated by $H_{sc}^{\rm RWA}$ if $A$ satisfies $A/\omega_0 \ll 1$. However, $H_{sc}^{\rm RWA}$ may be directly obtained as the limit of $H_q^{\rm RWA}$. It can be seen from Fig.~\ref{sc_converge} with $\lambda = 10^{-4}$ that the full quantum model agrees with the full semiclassical model and the qRWA agrees with the scRWA, despite the clearly discernible differences between the full and RWA dynamics. Looking at the trace distance plots, it is evident that both the trace distance between the full and RWA quantum state vectors and the trace distance between the corresponding reduced spin matrices tend toward the semiclassical result. This reflects the reduction of the joint spin-field dynamics to the tensor product form of Eq.~\eqref{H_limit}.

Figure~\ref{corr_coeffs_sc_converge} also shows that the convergence to the semiclassical limit as $\lambda$ is decreased is slower for smaller values of $A$. This may be understood by considering the relation between the Rabi frequency and the collapse timescale. The collapse time depends only on $\lambda$, while the Rabi frequency is given by $A = \lambda \alpha$. Therefore, for smaller values of $A$, fewer Rabi oscillations occur within the collapse time; over a fixed number of Rabi oscillations the effect of the collapse is more pronounced, resulting in a greater divergence between the quantum and semiclassical dynamics.

\section{Conclusions}\label{sec:conclusions}

The analysis presented here demonstrates that even when the field and spin are exactly resonant, the accepted statements of the regimes in which the qRWA is a good approximation to the full quantum Rabi model are far from capturing the full picture. The conventional definition of the ultrastrong coupling regime in terms of a fixed threshold $\lambda_c/\omega_0 = 0.1$ only applies for the lowest energy levels in the spectrum. Comparing several definitions of the coupling at which the full and RWA energies diverge, we find they consistently predict that $\lambda_c/\omega_0$ scales as $1/(2\sqrt{n})$ for larger $n$. This distinction becomes important for calculations of dynamics with large-amplitude fields. With a coherent state as the initial state of the field, the spectral validity condition together with the definition of the semiclassical limit imply that the validity of the RWA for dynamics should be solely determined by the scRWA validity condition $A = \lambda \abs{\alpha} \ll 1$. 

Testing this prediction necessitates having a definition for the dynamical validity of the RWA. Assessing the few existing measures and bounds against our numerical calculations of dynamics reveals that they are inadequate at best. As alternatives, we have considered the trace distance and the correlation between Fourier spectra as measures to study the dependence of the dynamical validity of the qRWA on $\lambda$ and $\alphamag$. Contradicting the prediction from spectral validity, the dynamical validity exhibits a complicated dependence on the two parameters independently, not just on their product. Convergence to the semiclassical limit, for both dynamics and validity measures, is only achieved in the limit of small coupling and large field amplitude and over suitably small timescales. This is consistent with the derivation of the semiclassical limit in Ref.~\cite{TwyIrish2022} and nicely illustrates how the limit is approached. We conclude that the spectral validity and dynamical validity should be delineated as distinct concepts.

In many ways, this work raises more questions than it answers. The trace distance and Fourier correlations display a surprisingly rich structure, and the origin of all their features is not yet fully understood. Neither is it established that these are the best choices of measures for the dynamical validity of either the qRWA or the scRWA. Identifying bounds on the dynamical validity that can be put to practical uses is another major open topic for future work. Once again, the quantum Rabi model throws up a surprising degree of complexity despite its apparent simplicity. This characteristic of the Rabi model, alongside its many and varied physical realisations, has fascinated physicists for the last 60 years. There is every reason to believe that it will continue to do so for many years to come.

\begin{backmatter}

\bmsection{Acknowledgments}
E.K.T. wishes to thank C. Knigge for stimulating discussions.

\bmsection{Disclosures}
The authors declare no conflicts of interest.


\bmsection{Data availability} Data underlying the results presented in this paper are not publicly available at this time but may be obtained from the authors upon reasonable request.

\end{backmatter}


\bibliography{RWA_validity}

\end{document}